\documentclass[fleqn,usenatbib]{mnras}
\usepackage{hyperref}
\usepackage{graphicx}
\usepackage{natbib}
\usepackage{float}
\usepackage{amsmath}

\usepackage{newtxtext,newtxmath}
\usepackage[T1]{fontenc}

\newcommand{\nt}{\textcolor{black}}
\newcommand{\nttwo}{\textcolor{black}}

\begin{document}

\title[Europa in the Post-Main Sequence]{Life after death: Europa in the evolving Habitable Zone of a Red Sun}


\author[Mullens et al. 2025]{
Elijah Mullens,$^{1}$\thanks{E-mail: eem85@cornell.edu}
Britney Schmidt,$^{1}$
Lisa Kaltenegger$^{1}$
Nikole K. Lewis$^{1}$
\\
$^{1}$ Department of Astronomy and Carl Sagan Institute, Cornell University, 122 Sciences Drive, Ithaca, NY 14853, USA\\
}

\date{Accepted 2025 May 09. Received 2025 May 09; in original form 2024 December 17}

\pubyear{2025}

\label{firstpage}
\pagerange{\pageref{firstpage}--\pageref{lastpage}}
\maketitle

\begin{abstract}
\noindent Most stars end their main-sequence (MS) lives by evolving through the red-giant and asymptotic-giant branches before ending as a quiescent, stable white dwarf. Therefore, it is imperative to model the post-MS as it relates to long-term stability of environments potentially suitable for life. Recent work has shown that gas giants can exist in the habitable zone (HZ) during the red giant phase and around a white dwarf remnant. Icy moons represent large reservoirs of water and will evolve through sublimation and melting when exposed to higher instellation, where the relatively lower surface gravity could lead to the rapid loss of all surface water. We model the surface evolution of Europa when initially exposed to habitable zone instellation \nt{in the red giant branch}. Modeling the diurnal and yearly flux variations on a 2D map we show \nt{that, due to Jupiter's increased albedo, the sub-Jovian hemisphere of Europa largely sublimates while only the anti-Jovian equatorial band sublimates. With the increasing instellation of the red giant branch, both hemispheres sublimate substantially.} We then model the evolution of a tenuous water-vapor atmosphere and show it is stable against atmospheric loss \nt{for at least 0.2 Gyr in the red giant branch habitable zone}. We then present \nt{three} ways to observe a sublimating Europan-liken exomoon and potential spectra. Extending the results of this work to different planets and moons could open up a new pathway by which life could persist beyond the death of a star.
\end{abstract}









\section{Introduction}

Stars depart from the main-sequence after burning a majority of the hydrogen in their core and evolve depending on their initial mass. Sun-like stars go through a red-giant and asymptotic-giant phase and end up as quiescent white dwarfs. More massive stars experience a supernova and end up as either neutron stars or black holes. Red giants, white dwarfs, neutron stars, and other phases are collectively termed post-main sequence stars \citep[for a thorough review of post-main sequence stellar evolution, see][]{Iben2013Vol1, Iben2013Vol2}. 

Exoplanets are planets orbiting stars outside of our solar system. Exoplanetary science's history is inseparably tied to post-main sequence stars. The first exoplanets were detected in 1992 around a post-main sequence star: the pulsar PSR1257 + 12 \citep{Pulsar_Planet_Discovery}. Although pulsar planets remain rare as the field has advanced, evidence for planetary bodies around white dwarfs via white dwarf pollution dates back even earlier to the 1910s \citep{white_dwarf_1910}. It has been extrapolated that over 90\% of all observed exoplanets orbit a star that will end up as a white dwarf and that eventually, white dwarfs will be the most common star in the Milky Way \citep{Veras_Post_MS_Review}. Therefore, understanding how planetary systems evolve around red giant stars and white dwarfs is crucial when considering the fate of the majority of exoplanets, and \nt{is} interesting for the study of our own solar system.

In particular, white dwarf planetary systems provide an exciting new population of potentially habitable planets that can be characterized by current space-based observatories. White dwarfs have been found to have water-rich material orbiting them \citep[e.g.,][]{White_Dwarf_Water_Rich} and remain stable on the order of the age of the universe \citep[e.g.,][]{White_Dwarf_Formation}. Exoplanets orbiting white dwarfs make fantastic observation targets for detectability and characterization due to the host star's quiescence and the large radius ratio between planet and star \citep{White_Dwarf_Jupiter}. Though white dwarf systems provide great characterization targets, planets have to survive up to that point and the evolution from a main sequence star to a white dwarf is drastic. In order to model planet survivability one must account for a variety of time-dependent factors during the red giant branch such as the host star's changing luminosity, orbital expansion and decay of planetary bodies (i.e., the increase of a planet's semi-major axis due to stellar mass loss or decrease due to tidal interactions), and the hot and dense stellar winds from the host star accreting onto or ablating planetary bodies, just to name a few \citep[for a more complete review, see][]{Veras_Post_MS_Review}. Inner terrestrial bodies, in particular, are prone to ablation and destruction \citep[e.g.,][]{RGB_solar_system_terresterial_body_evolution}. The recent microlensing event KMT-2020-BLG-0414 \citep{Zhang2024} confirmed that Earth-like planets can survive to the white dwarf phase, but its unknown how the surviving planet's surface evolved. The inevitability of inner terrestrial body destruction or surface evolution makes it difficult to imagine rocky bodies with surface water existing in a red giant's or white dwarf's habitable zone (HZ). However, the survivability of outer gas giants is much more likely \citep[e.g.,][]{RGB_orbital_evolution}. The discovery and characterization of gas giant WD1856+534b \citep[][MacDonald et al. 2024 (submitted)]{WD1856b_Discovery} is especially thought-provoking since it orbits in the habitable zone of a white dwarf and has been the target of James Webb Space Telescope (JWST) transit observations.

\citet[Fig 5 \& Table 3]{Jupiter_in_HZ} study in detail the post-MS evolution of the solar system and show that Jupiter will exist in the habitable zone of the red giant branch (RGB) phase for $\sim$0.37 Gyr, after which it will migrate outwards to $\sim$19 AU. Jupiter can then possibly migrate inwards via Lidov-Kozai migration, interacting with the planetary nebula \citep{LidovKozai1,LidovKozai2} in order to become similar to WD1856+534-b and orbit within or near the habitable zone of a white dwarf. This begs us to question the habitability of terrestrial and icy moons orbiting Jupiter-sized planets during and after post-main sequence evolution. Icy moons represent large reservoirs of terrestrial H$_2$O in our solar system \citep[e.g.,][]{Schmidt2020} and could provide an interesting class of potentially habitable planets that retain surface or subsurface liquid water intermediately during the red giant branch and long-term in the white dwarf phase.

Europa is an icy moon orbiting Jupiter that has a liquid ocean under a thick ice shell. When the Jovian system is exposed to increasing instellation during the red giant branch the surface ice of Europa will begin to evolve through sublimation \citep[e.g.,][]{Hobley2018}. Due to Europa's low gravity it is expected that a portion of the sublimated atmosphere would be rapidly lost and that surface water on Europa would be unstable on long timescales \citep[e.g.,][]{Szalay2024}. 

In this work, we model two distinct snapshots of Europa's surface water evolution in the habitable zone of a post-MS star: the initial melting and sublimation of Europa's icy surface, and the atmospheric evolution of Europa once it has sublimated a substantial secondary atmosphere. In \S \ref{sec:Melting} we leverage a 2D surface evolution scheme to model the initial sublimation of ice on Europa's surface \nt{when the Jupiter-Europa system first enters the red giant branch habitable zone, as well as it when it receives Earth-like instellation,} as it relates to diurnal flux changes (due to Europa's unique orbital configuration), yearly flux changes (due to Jupiter's obliquity and orbital eccentricity), and \nt{$\sim$Gyr changes in the incident stellar flux and Jupiter's albedo.} In \S \ref{sec:Atmosphere} we employ first-order considerations of mass loss and photolysis to model how the tenuous atmosphere of Europa could evolve once substantial sublimation has occurred. After showing that Europa readily forms and maintains an atmosphere \nt{for at least 0.2 Gyr}, we then explore the potential observability of Europa with a sublimated atmosphere during the red giant and white dwarf phases. In \S \ref{sec:Exomoon} we present \nt{three} methods by which to detect Europan-liken exomoons with an evolved water-rich atmosphere and potentially characterize them. In \S \ref{sec:Discussion} we discuss the results and propose additional considerations to model this system with more complexity. Finally, we conclude in \S \ref{sec:Conclusions}. 


\section{\nt{Sublimating} Europa}\label{sec:Melting}

In order to test whether or not Europa's surface substantially sublimates we model the diurnal and yearly variations of absorbed flux on the surface of Europa as it varies with latitude and longitude \nt{for two points in time during the red giant branch}. We then apply a simple Newtonian cooling scheme \nt{with phase changes included} to compute  of the surface temperature and phase of surface water over the \nt{course of single Europan day once the model has converged to a steady-state solution.} 


\begin{figure*}
     \centering
         \includegraphics[width=1.0\linewidth]{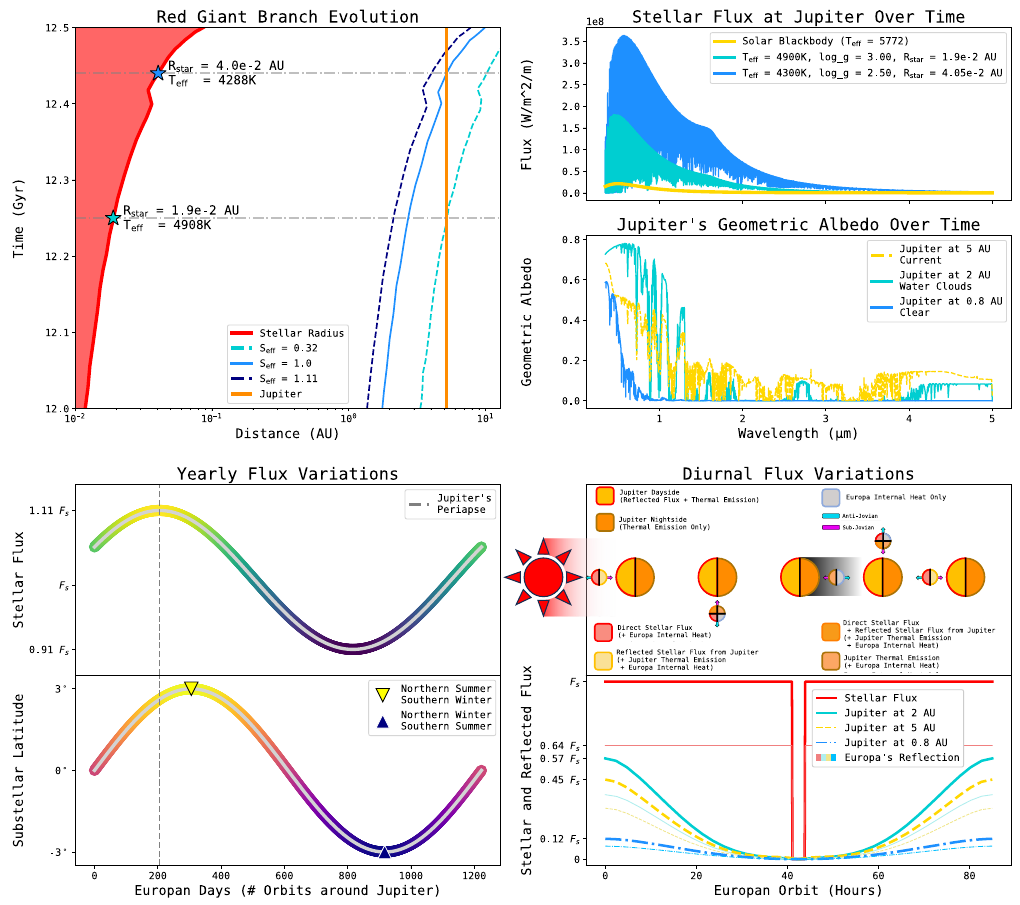}
         \caption{\nt{Model assumptions for stellar and reflected flux variations over different timescales. Top: The Sun's red giant branch evolution and incident flux at Jupiter over time (Left). Stellar flux at Jupiter (\texttt{PHOENIX} stellar models \citep{Husser2013}) and geometric albedo (a 0$^\circ$ phase) of Jupiter \citep{Cahoy2010} when it first enters the red giant branch habitable zone (S$_{\mathrm{eff}}$ = 0.32, turquoise) and when it receives Earth-like instellation (S$_{\mathrm{eff}}$ = 1.0, blue) (Right). Stellar spectra and geometric albedo are convolved to derive the phase-dependent bond albedo of Jupiter. Bottom:} Yearly stellar flux variations due to Jupiter's eccentricity and obliquity (Left), and the diurnal stellar and reflected flux variations due to Europa's orbit around Jupiter (Right). We find that Europa's northern hemispheric solstice (substellar latitude = 3 $^\circ$) occurs near Jupiter's periastron. For the diurnal flux, we include the direct stellar flux absorbing on the surface of Europa, the phase dependent reflected flux from Jupiter, and factor in Jupiter eclipsing the stellar flux for a portion of the orbit. \nt{Over-plotted is the amount of flux Europa intrinsically reflects away.} \nttwo{The shading scheme for the Diurnal Flux Variations plot (bottom right, top panel) was adapted from \citet{HellerBarnes2013}.}}
     \label{fig:Yearly-Diurnal-Flux}
\end{figure*}

\subsection{\nt{Evolution during the Red Giant Branch}}

\nt{We first consider the incident stellar flux at the Jupiter-Europa system during two points in the red giant branch. The first is when the Jupiter-Europa system enters the habitable zone of the red giant branch at $\sim$12.25 Gyr. At this point, the system receives S$_{\mathrm{eff}}$ = 0.32 (as defined in \citet{Jupiter_in_HZ} this is the incident flux when Mars still had water on its surface, corresponding to an incident flux of 439.36 W/m$^2$). We also consider the point when the Jupiter-Europa system receives  S$_{\mathrm{eff}}$ = 1.0 (the incident flux at modern-day Earth, corresponding to an incident flux of 1373 W/m$^2$) at $\sim$12.45 Gyr. These two points in time are highlighted in Figure \ref{fig:Yearly-Diurnal-Flux} (top left).}

\nt{Given the tidally locked configuration of Europa, it is essential to model the phase-dependent amount of of reflected light from Jupiter at these two points in time in the red giant branch. As the incident flux at Jupiter increases over time, it's atmospheric composition and resulting geometric albedo will change. \citet{Cahoy2010} model Jupiter's resultant phase-dependent geometric albedo when receiving incident flux of a Sun-like star at 0.8 AU, 2 AU, 5 AU, and 10 AU. Currently, Jupiter's geometric albedo (at 5 AU) is due to its atmospheric composition and top-most cloud layers (ammonia ice and hazes). When placed at 2 AU, Jupiter's top-most cloud layer becomes water clouds which have a high albedo. At 0.8 AU, these water clouds dissipate and the albedo becomes dominated by gas-scattering and absorption (resulting in an overall lower geometric albedo). We take the 2 AU model to be indicative of Jupiter's geometric albedo when the Jupiter-Europa system first enters the red giant branch habitable zone, and the 0.8 AU model to be when it receives an incident flux of S$_{\mathrm{eff}}$ = 1.0. The geometric albedo of Jupiter at 0$^\circ$ phase we utilize can be seen in Figure \ref{fig:Yearly-Diurnal-Flux} (top right, bottom panel).} 

\nt{During the course of the red giant branch, the stellar spectrum evolves. Notably, the star becomes more luminous (due to its radius growing) and it's peak wavelength shifts to more red wavelengths. We have taken high-resolution \texttt{PHOENIX} stellar models \citep{Husser2013} for the two points in time we investigate. We then compute a phase dependent bond albedo by normalizing the phase dependent geometric albedo by the stellar spectrum (see Eq 6 in \citet{Cahoy2010}); T$_{\mathrm{eff}}$ = 4900K for 2 AU and T$_{\mathrm{eff}}$ = 4300K for 0.8 AU. The phase dependent bond albedo of Jupiter can be seen in Figure \ref{fig:Yearly-Diurnal-Flux} (bottom right, bottom panel).}

\subsection{Diurnal and Yearly Flux Variations}

Europa is tidally locked to Jupiter, meaning that one hemisphere always faces away from Jupiter (anti-Jovian hemisphere) and one hemisphere always faces towards Jupiter (sub-Jovian hemisphere). Given the unique orbital configuration of synchronously tidally-locked moons, it is expected that the anti- and sub-Jovian hemispheres will experience different flux variations during the course of a single Europan orbit. Due to this asymmetry, we decided to model the system using a latitude-longitude mapping. We split the total absorbed flux into four different components: absorbed stellar flux $F_s$, absorbed reflected flux from Jupiter $F_r$, absorbed thermal emission from Jupiter $F_t$, and internal outgoing flux due to tidal forces on Europa $F_q$. 

The absorbed stellar and reflected flux both have yearly and diurnal variations. The yearly variation comes from Jupiter's eccentric orbit (ecc = 0.0489) as well as Jupiter's obliquity ($\sim$3$^\circ$). Over the course of a Jovian year, Jupiter's orbital distance \nt{varies} from 4.9506 AU to 5.4570 AU, resulting in an incident stellar flux that \nt{goes} from 0.909 to 1.105 its base value. Assuming an incident stellar flux of \nt{439.36 W/m$^2$ (1373 W/m$^2$) [corresponding to S${\mathrm{eff}}$ = 0.32 (S${\mathrm{eff}}$ = 1.0)], the flux swings from 399.38 W/m$^2$ (1247.97 W/m$^2$) to 485.49 W/m$^2$ (1517.81 W/m$^2$) over the period of Jupiter's orbit (11.86 years, or $\sim$1223 Europan orbits).} The obliquity of Jupiter determines the substellar latitude, i.e. the latitude at which a planetary body receives the most direct incident flux. For the Jovian system, this \nt{can be} $\pm$ $\sim$3$^\circ$ from the equator. Seasonal effects have been shown to be miniscule on Jupiter but are potentially important for its moons \citep[e.g.,][]{Ashkenazy2016, Orton2023}. For Europa, \citet{Ashkenazy2016} showed that seasonal variations can increase polar surface temperatures by $\sim$10K. We take the average obliquity of Europa to be 3$^\circ$ following \citet{Ashkenazy2016}. We utilize the Jupiter Ephemeris Generator 3.0 \citep{Chanover2022} to constrain the substellar latitude during Jupiter's periapse (January 21, 2023) to $\sim$2.61$^\circ$ (increasing to 3$^\circ$). This means that Europa experiences a northern hemisphere summer close to Jupiter's periapsis. Both the instantaneous incident stellar flux and substellar latitude are then taken as constants during the course of a single Europan day and are shown in Figure \ref{fig:Yearly-Diurnal-Flux} (bottom left).

During the course of Europa's 85-hour-long orbital period, the substellar longitude moves westward at a rate of 4.325$\frac{\circ}{hr}$. All incident stellar flux will be blocked during the Jupiter eclipse, affecting the sub-Jovian hemisphere. \nt{For all orbits we assume that Europa is near equinox (where eclipse path-length is equal to Jupiter's diameter) since the maximum difference in path-length is 0.1\% Jupiter's diameter at Europa's maximum inclination (0.47$^\circ$).} Due to tidal locking and the Laplace resonance between Io, Europa, and Ganymede the orbit of Europa is prevented from being completely circularized \citep[see reviews][]{MurrayDermott1999, Planetary_Astrobio_Europa}. However, the eccentricity of Europa's orbit is small, $\sim$0.009 \citep{Europa_orbital_data}, so we assume a circular orbit approximation. We derive the amount of time Europa experiences Jupiter eclipsing the Sun by assuming a circular orbit of Europa around Jupiter where the orbital velocity is given by: 
\begin{equation}
v_p = \frac{2 \pi a}{P} = 1.37e4 \: \frac{m}{s}
\end{equation}
where the orbital radius is $a$ = 6.67e8 m and the orbital period is $P$ = 85 hours. The diameter of Jupiter is $D_j$ = 13.982e7 m. Therefore, Europa is eclipsed by Jupiter for 2.8 hours out of the 85-hour orbit. This is consistent with \citet{Ashkenazy2016} where more complex eclipse modeling was utilized. \nt{ We additionally assume the eclipse occurs instantaneously due to its egress duration ($\sim$4 minutes) being shorter than our model time resolution (6 minutes).}

A latitude-longitude grid of a spherical body was modeled adapting the methodology developed by \citet{Eccentric_Planet_Radiative_Timescale} for a diurnal stellar flux. The absorbed stellar flux, $F_{s}$, at a given latitude-longitude is given by 
\begin{equation}
F_{s}(t, \theta, \Phi) = (1-A) \:\; F_{inc,s}(t) \:\; \mathrm{sin}(\theta+\theta_s) \:\; \mathrm{max}(\mathrm{cos}(\Phi(t)),0)
\end{equation}
where $A$ is the Bond albedo of Europa, $F_{inc,s}(t)$ is the incident flux at time $t$, $\theta$ is the latitude angle ($0^{\circ}$ at the north pole and $180^{\circ}$ at the south pole), $\theta_s$ is the substellar latitude, and $\Phi(t)$ is a point's longitudinal angle from the substellar longitude. The sine factor ensures that the maximum absorbed stellar flux occurs at the substellar latitude. To guarantee that the polar regions receive zero flux when the substellar latitude is nonzero, we take the sine factor as zero where it goes negative. $\Phi(t)$ is defined as $0^{\circ}$ at the substellar longitude, $90^{\circ}$ at the sunset longitude, $180^{\circ}$ at the midnight longitude, and $-90^{\circ}$ at the sunrise longitude. 

\nt{In the previous section, we modeled the phase-dependent bond albedo of Jupiter to model its reflected light component.} Given Europa's relative size and low inclination to Jupiter, we model this as an incident flux independent of latitude and longitude on the sub-Jovian hemisphere. The incident diurnal stellar and reflected flux is shown in Figure \ref{fig:Yearly-Diurnal-Flux} (bottom left, bottom panel). Throughout the entire Europan orbit, the sub-Jovian hemisphere is receiving thermal emission from Jupiter, which is taken as a constant 0.176 W/m$^2$, and the entire surface is receiving a flux contribution is the constant tidal heating flux \nt{F$_\mathrm{q}$} = 0.05 W/m$^2$ \citep{Ashkenazy2016}. \nt{We opt to use modern-day values for Jupiter's thermal emission and Europa's tidal heating flux since Jupiter is no longer rapidly cooling \citep{Howard2024}, and there are many unknowns to how Jupiter and Europa's internal heat will evolve during the red giant branch.} Note that neither the thermal emission nor internal heating term utilizes Europa's albedo when considering absorbed flux. The total absorbed flux is given by: 

\begin{align}
F_{abs, tot}(t, \theta, \Phi) &= F_{s}(t, \theta, \Phi) \\
&+ (1 - A)*F_{r}(t, \theta, \Phi)\\
&+ F_{t}(t, \theta, \Phi)  + F_{q} \; .
\end{align}

\subsection{2D Temperature Model}

\begin{figure*}
     \centering
         \includegraphics[width=1.0\linewidth]{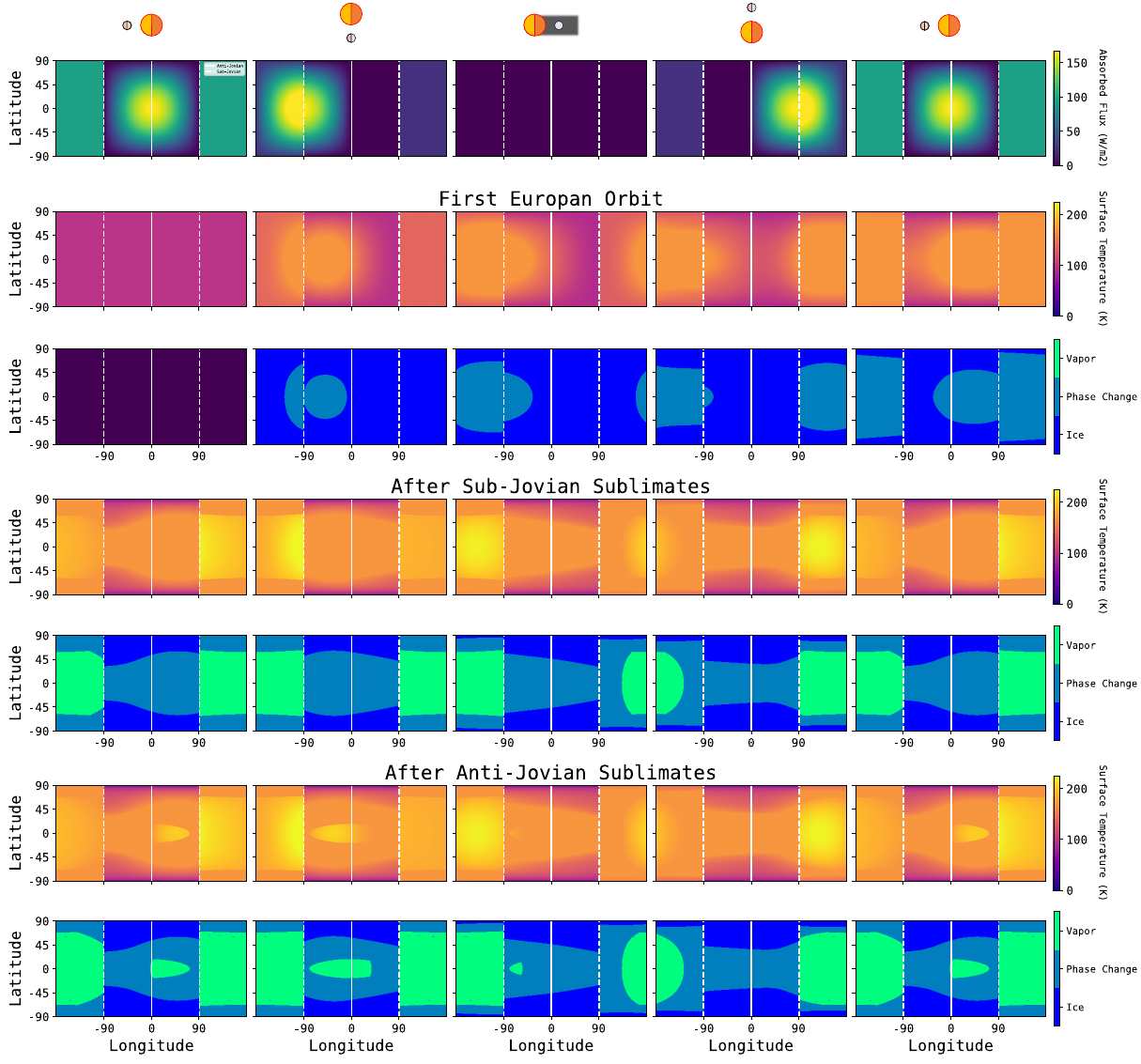}
     \caption{\nt{Snapshots of the latitude-longitude simulation of Europa's surface evolution when it first enters the red giant branch habitable zone (12.25 Gyr, S$_{\mathrm{eff}}$ = 0.32, Jupiter at 2 AU \citet{Cahoy2010} albedo). First three rows show snapshots of the absorbed flux, surface temperature, and phase of Europa's surface during the first integration of the simulation. The substellar point contributes most of the flux near the equator as Europa orbits Jupiter, and the reflected flux from Jupiter keeps the sub-Jovian hemisphere warmer than the anti-Jovian hemisphere. Much of the surface readily reaches the sublimation point of 170K and begins to undergo the phase change. After $\sim$6 Jupiter years, the mid-latitudes and equator of the sub-Jovian hemisphere complete the phase change and sublimate, allowing that hemisphere to heat above 170K. After $\sim$15 Jupiter years, the equator of the anti-Jobian hemisphere completes the phase change and sublimates. Yearly and daily surface temperature variations after the simulation has reached steady state are shown in Figure \ref{fig:2AU-temperatures}.}}
     \label{fig:2AU-Snapshots}
\end{figure*}

\begin{figure*}
     \centering
         \includegraphics[width=1.0\linewidth]{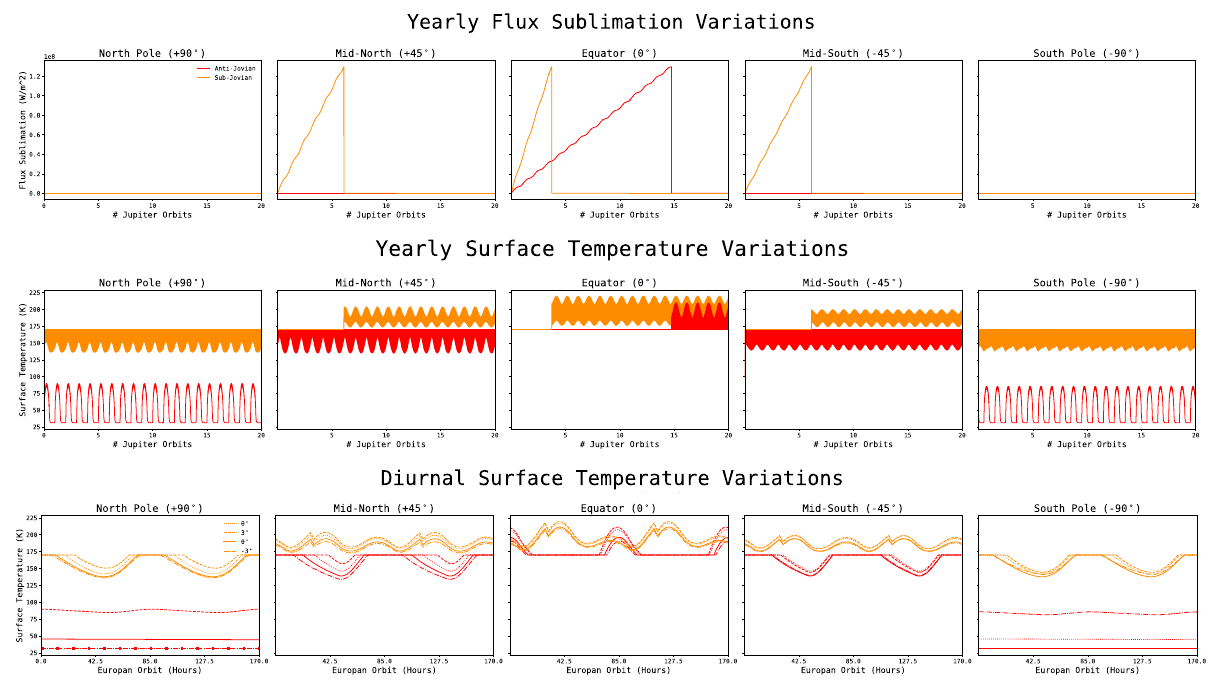}
         \caption{\nt{Flux Sublimation and Surface Temperature for specific latitudes (+90$^\circ$,+45$^\circ$,0$^\circ$,-45$^\circ$, and -90$^\circ$ at the anti-Jovian and sub-Jovian hemispheres) for the Europa surface evolution simulation when it first enters the red giant branch habitable zone (12.25 Gyr, S$_{\mathrm{eff}}$ = 0.32, Jupiter at 2 AU \citet{Cahoy2010} albedo). Top Row: Once the sub-Jovian mid-latitudes and both hemisphere's equatorial regions reach the sublimation temperature (170K), they begin a phase change (from ice to vapor) where each pixel undergoing the phase change must accumulate more flux with each Europan orbit than they thermally emit (to eventually equal the latent heat). Due to both stellar and reflected flux contributions, the sub-Jobvian hemisphere sublimates after $\sim$6 Jupiter years while the anti-Jovian hemisphere sublimates after $\sim$15 Jupiter years. Middle Row: Year-long seasonal variations in surface temperature. Due to Jupiter's eccentric orbit there are variations in the peak surface temperature, and the northern latitudes display more extreme seasonal differences due to Europa experiencing northern summer during Jupiter's periapse. Bottom Row: Surface temperature variations for two Europan orbits (at four different times in the Jupiter year, where the first 0$^\circ$ represents the substellar latitude coming from -3$^\circ$) when the simulation has reached a steady state. The sub-Jovian hemisphere's temperature is much more dynamic due to the eclipse keeping it from reaching it's peak temperature, as well as the reflected flux from Jupiter keeping it warmer than the anti-Jovian hemisphere. The max surface temperature of the sub-Jovian (anti-Jovian) hemisphere is 220K (212K).}}
        \label{fig:2AU-temperatures}
\end{figure*}

We implement a simple Newtonian cooling method to convert absorbed flux to an approximate surface temperature. We first compute the surface equilibrium temperature assuming that the layer above the surface has zero heat capacity via:
\begin{equation}
T_{eq, surface}(t, \theta, \Phi) = (\frac{F_{abs, tot}}{\epsilon\:\sigma})^{1/4}
\end{equation}
where $\epsilon$ is the emissitivity of Europa (= 0.94). The equilibrium surface temperature determines whether or not the true surface temperature will heat up or cool down. We initialize a latitude-longitude grid with each pixel set at 98K. We then time step Europa's orbit every six minutes and compute the surface equilibrium temperature for each point given the absorbed flux. The change in temperature at each point is given by: 
\begin{equation}
dT/dt(t, \theta, \Phi) = \begin{cases} 
\frac{F_{abs,tot}-\epsilon\sigma T_c^4}{c_h} & \text{if } T_c \neq 170K \\
0 & \text{if } T_c = 170K \\
\end{cases}
\end{equation}
where 
\begin{equation}
c_h = \rho C_p H
\end{equation}
and $\rho$ is the mass density (917 kg/m3 for ice, 0.6 kg/m3 for water vapor), $C_p$ is the specific heat capacity (2.108 kJ/kgK for ice, 1.996 kJ/kgK for water vapor) and H is the thickness of the parcel being heated up. Europa's surface ice has a lower thermal inertia than ice found on Earth's surface \citep[e.g.,][]{Spencer1999, Rathbun2010}. \citet{Ashkenazy2016} modeled the surface temperature oscillations of Europa and found the thermal depth at the equator to be $\sim$0.05 m. We take H to be 0.05 m. If the current temperature ($T_c$) is lower than the equilibrium temperature the temperature at that latitude-longitude increases by dT/dt (and vice versa). 

We take the phase transition between ice and vapor to occur at \nt{170K} for the current negligible surface pressure of Europa (1e-12 bars), \nt{where the sublimation temperature at decreasing pressures is nominally between 150K-200K \citep{Mortimer2018,Yue2023}. When a pixel reaches the sublimation temperature, a phase change starts to occur. During the phase change, the temperature doesn't change and the pixel must accumulate energy equal to the latent heat in order to change phases:
\begin{equation}
Q = \begin{cases}
\sum (F_{abs,tot}(t)-\epsilon\sigma (170K)^4)\times Area & \text{if } \mathrm{Sublimating} \\
\sum (\epsilon\sigma (170K)^4 - F_{abs,tot}(t))\times Area & \text{if } \mathrm{Condensing} \\
\end{cases}
\end{equation}
where $Q = mL$ = 1.2e8 J is the latent heat, $m$ is the mass of a single pixel ($Area$ is the area of a single pixel) and $L$ = 2830 kJ/kg is the latent heat of ice sublimation \citep{Feistel2007,Yue2023}. In other words, a pixel undergoing a phase change from ice to vapor must accumulate more flux than it thermally emits each orbit of Europa (denoted as Flux Sublimation in our results). For condensation, the pixel must lose more flux due to thermal emission than it accumulates each orbit to complete a phase change. If the sum reaches 0, the phase change will stop and the pixel will be allowed to thermally evolve again without undergoing the phase change. After a sublimation phase-change is completed, the specific heat capacity of water vapor is used in lieu of water ice, and vice versa. In our model, we assume that the entire surface of Europa is composed of water ice. While recent observations of Europa's surface have revealed salts and carbon-bearing material, the bulk surface composition is primarily water ice that controls its surface evolution and processes \citep{King2022,Cartwright2025}.}

\subsection{Results and Discussion}
We utilized the diurnal and yearly flux variation model alongside a 2D Newtonian cooling method to explore the surface evolution of Europa \nt{when it first enters the red giant branch habitable zone (12.25 Gyr, S$_{\mathrm{eff}}$ = 0.32, Jupiter at 2 AU \citet{Cahoy2010} albedo) and when it receives Earth-like instellation (12.45 Gyr, S$_{\mathrm{eff}}$ = 1.0, Jupiter at 0.8 AU \citet{Cahoy2010} albedo). The model time steps every 6 minutes (850 time steps per Europan orbit), where we begin the models at a point in Jupiter's orbit where the substellar latitude is 0$^\circ$ (increasing to 3$^\circ$) and run the models until they reach a steady-state.}

\nt{We will first focus on the results of the simulation when Europa first enters the red giant branch habitable zone. Figures \ref{fig:2AU-Snapshots} and \ref{fig:2AU-temperatures} displays the results of the combined modeling efforts of diurnal and yearly surface temperature variations. The top row of Figure \ref{fig:2AU-Snapshots} shows the absorbed flux due to incident stellar flux (which shifts across Europa as it orbits Jupiter), reflected light from Jupiter incident on the sub-Jovian hemisphere (which is a maximum at 0$^\circ$ phase in the first panel, and decreases throughout the orbit), and Jupiter eclipsing the Sun. During the first orbit of Europa, the equatorial regions and mid-latitudes of the sub-Jovian hemisphere rapidly reach the sublimation temperature (170K, Figure \ref{fig:2AU-Snapshots}, second and third rows) and begins a phase change. Due to the low thermal inertia of the surface ice, as well as the small thermal depth, the surface of Europa heats and cools relatively fast compared to its orbital period. However, in the mid-latitudes of the sub-Jovian hemisphere, and the equatorial region of both hemispheres, more flux is absorbed than thermally emitted with each Europan orbit (see Figure \ref{fig:2AU-temperatures}, top row), allowing the phase-change to eventually complete. After $\sim$6 Jupiter years, regions of the sub-Jovian hemisphere convert to a vapor and can begin heating past 170K (Figure \ref{fig:2AU-Snapshots}, third and fourth rows). After $\sim$15 Jupiter years the anti-Jovian equator also converts to a vapor (Figure \ref{fig:2AU-Snapshots}, fifth and sixth rows). We allow the simulation to run for a total of 20 Jupiter years in order to find steady-state yearly and diurnal surface temperature patterns.}

\nt{The second row of Figure \ref{fig:2AU-temperatures} displays the surface temperature of Europa at five different latitudes on both hemispheres for the entirety of the simulation (note that due to the latitude-longitude mapping, $\pm 90^{\circ}$ represents the top and bottom rows of the models and is therefore representative of the polar region as a whole). We find that the northern latitudes experience more drastic fluctuations due to the Jupiter's periapse aligning with northern summer (+3$^\circ$), and likewise Jupiter's apoapse with northern winter (-3$^\circ$). Once the simulation has reached a steady state we take diurnal surface temperatures (over two Europan orbits) for four different times of Jupiter's year (namely, the equinoxes  $0^{\circ}$ [coming from $-3^\circ$ first] and solstices $\pm 3^{\circ}$).
}

\nt{Starting with the poles, the anti-Jovian polar surface temperatures display an expected behavior; namely, the temperatures remain low except during solstices when they can receive direct stellar flux (after which, they thermally cool). The sub-Jovian polar regions receive more absorbed flux due to Jupiter's reflected flux. These polar regions reach the sublimation temperature (and start to change phases) each Europan orbit, but do not accumulate enough flux to remain in that state, and eventually cool below the sublimation temperature. Therefore, the poles of both hemispheres remain in the ice phase.}

\nt{The mid-latitudes of the anti-Jovian hemisphere, similar to the polar region of the sub-Jovian hemisphere, reach the sublimation temperature each Europan orbit but do not proceed with the phase change and eventually cool below the sublimation temperature. The sub-Jovian mid-latitudinal region, on the other hand, accumulates enough flux to sublimate after $\sim$6 years. Both the anti-Jovian and sub-Jovian equatorial regions sublimate, with the anti-Jovian hemisphere taking $\sim$15 Jupiter years, and the sub-Jovian hemisphere taking $\sim$4 Jupiter years. Water vapor heats and cools rapidly to changes in absorbed flux over the course of a single Europan orbit. While receiving direct stellar flux, the anti-Jovian hemisphere reaches a peak temperature of 212K while the sub-Jovian hemisphere reaches a peak temperature of 220K. Each orbit, the anti-Jovian equatorial region rapidly cools back down to the sublimation temperature after the susbtellar point shifts away and begins to condense, but due to the latent heat release of the phase change, remains a vapor. For the sub-Jovian hemisphere, the eclipse somewhat regulates the surface temperature by keeping it from reaching the expected peak temperature when the surface would receive the most direct stellar flux. The combined effect of Jupiter's reflected light and absorbed stellar flux keeps the sub-Jovian hemisphere `warmer' compared to the anti-Jovian hemisphere, however the eclipse could interestingly regulate the surface temperature and combat atmospheric escape by keeping the surface temperature below a threshold temperature.}

In addition to anti- and sub-Jovian hemispheric differences, the north and south hemispheres experience differences due to seasonal variations. During a solstice, the temperature in the polar regions can be upwards of $\sim$20-50K hotter than nominal temperatures, and the mid-latitudes can experience temperature shifts that cause the peak surface temperatures to be $\sim$25K hotter. Additionally, as noted previously the northern hemisphere of Europa experiences a solstice near Jupiter's periapse. \nt{Therefore, seasonal differences are more extreme in norther latitudes}.

\nt{We also ran the simulation when the Jupiter-Europa system recieves full Earth-like instellation (12.45 Gyr, S$_{\mathrm{eff}}$ = 1.0, Jupiter at 0.8 AU \citet{Cahoy2010} albedo), with results found in Figures \ref{fig:08AU-snapshots} and \ref{fig:08AU-temperatures}. The main difference in this simulation is that the incident stellar flux is higher (1373 W/m$^2$ vs. 439 W/m$^2$) and the bond albedo of Jupiter is much lower due to cloud dissipation (peak bond albedo is 0.12 vs. 0.57). This results in the entire mid-latitudinal band of both hemispheres sublimating, with symmetric surface temperatures oscillations for both hemispheres. One notable difference is that the anti-Jovian hemisphere now reaches a higher maximum temperature (314K) than the sub-Jovian hemisphere, due to Jupiter's eclipse of the Sun driving cooling of the sub-Jovian hemisphere (306K).}

\nt{The modeling above shows that when the Jupiter-Europa system first enters the habitable zone of the red giant branch, the entire equatorial region of Europa's surface sublimates due to its small thermal inertia and thermal depth, with the mid-latitudinal region of the sub-Jovian hemisphere also sublimating. Taking into account 2D and diurnal/seasonal effects, we show that there exist interesting asymmetries between the north and south and anti- and sub-Jovian hemispheres. Namely, the anti-Jovian surface temperature oscillates solely due to incident stellar flux as the substellar point advances around the moon while the sub-Jovian surface temperature remains at a higher temperature due to incident stellar flux for half the Europan orbit and reflected flux from Jupiter (with bright water clouds) for the other half. However, the temperature of the sub-Jovian hemisphere is also slightly regulated by Jupiter eclipsing the Sun. The northern hemisphere experiences more drastic seasonal changes due to solstices aligning with minimum and maximum distances in Jupiter's eccentric orbit. A more full exploration of the surface evolution of Europa would require a 3D global climate model (GCM) with recirculation to expand on the patterns and potential vapor redistribution found in the 2D model. However, our simplified model is able to show that water vapor is produced and remains in a vapor phase in the equatorial region of the entire moon. Knowing that a vapor atmosphere forms, we explore the stability and evolution of this tenuous atmosphere in the next section.}

\section{Europan Atmosphere}\label{sec:Atmosphere}

As a small-mass, low-gravity object Europa's tenuous atmosphere could be susceptible to mass-loss mechanisms such as hydrodynamic escape and Jeans escape. We compute first order approximations of mass-loss in order to estimate the stability of surface water on Europa \nt{at temperatures Europa's surface and atmosphere might experience while in the red giant branch habitable zone}. With these mass-loss rates, we predict that Europa, without additional considerations, could retain a water-vapor and ice interface on its surface for long timescales. We then run a simple photochemical model assuming a substantial water vapor atmosphere to generate photochemical ionization rates that could result in rapid water loss.

\begin{figure*}
     \centering
         \includegraphics[width=1.0\linewidth]{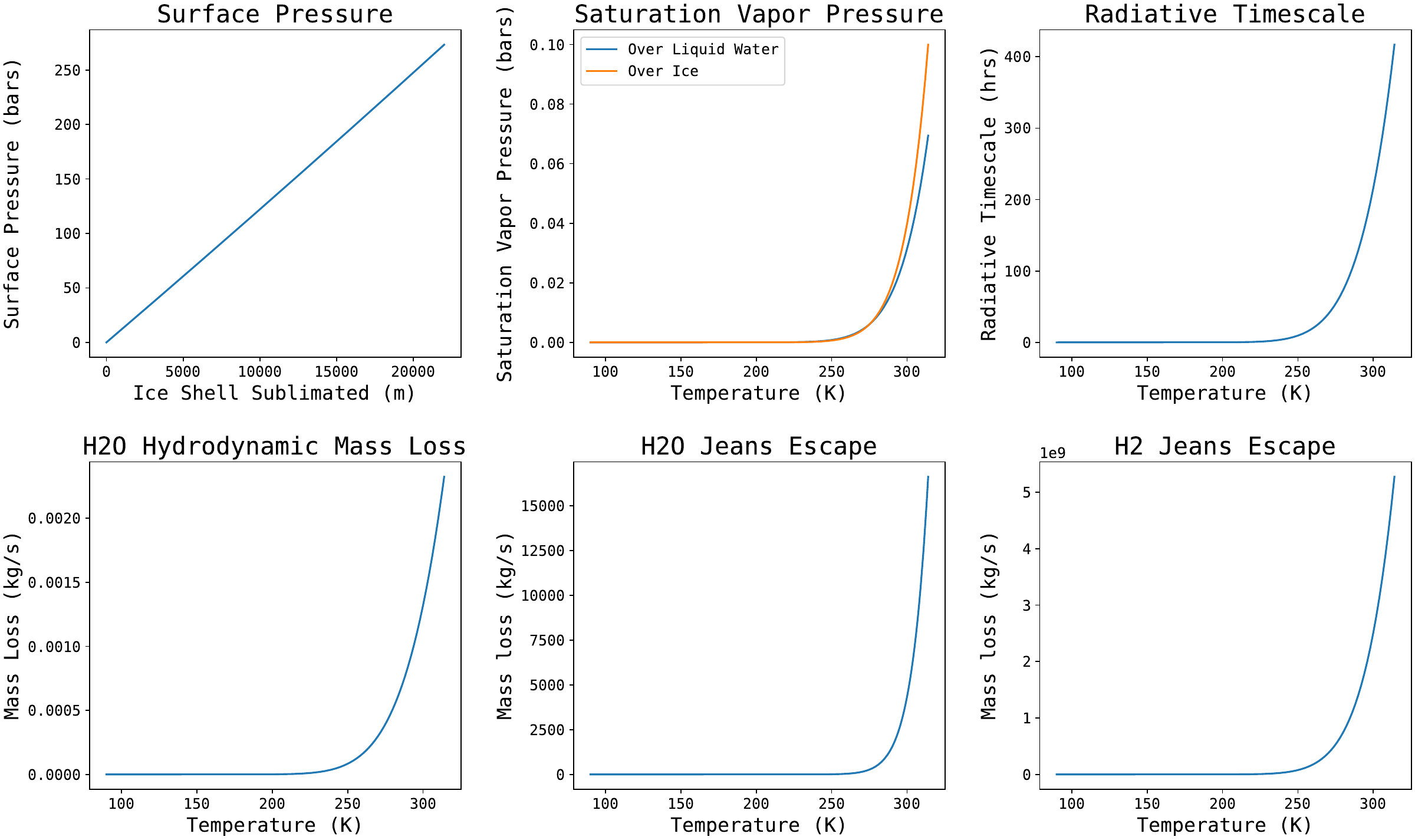}
     \caption{First panel displays the predicted surface pressure as it varies with the depth of the ice shell sublimated. The surface pressure quickly grows to unphysical levels. The second panel displays the saturation vapor pressure at a vapor-ice and vapor-liquid interface, given by the Aden-Buck equation, as it varies with temperature. Utilizing the saturation vapor pressure, the radiative timescale and mass loss rates are derived as they vary with temperature. At around 250K there is a sharp uptick in the radiative timescale and mass loss rates. This means that the atmosphere could reach a runaway phase or lose a substantial amount of water. However, \nt{temperatures this high only occur once the Jupiter-Europa system reaches Earth-like instellation at $\sim$12.45 Gyr.}}
     \label{fig:Timescales}
\end{figure*}


\subsection{Surface Pressure Evolution}

The modeling in \S \ref{sec:Melting} showed that a fractional depth of Europa's sublimates relatively quickly. It would then follow that, without condensation, the surface would continue to sublimate and begin to build up an atmosphere overlaying the surface ice. As the atmosphere begins to build up the surface pressure will increase. We assume that the atmosphere now represents a global property of Europa and that the amount of atmosphere produced is directly proportional to the depth of the ice shell being sublimated. Taking $z$ to be the ice shell thickness being sublimated, the volume of the ice shell is given by
\begin{equation}
\label{eqn:shell_volume}
V_{shell} = \frac{4}{3} \pi (R_{Europa}^3 - R_{Ocean}^3)
\end{equation}
where $R_{Ocean} = R_{Europa} - z$. The volume of ice is converted to mass using ice density $\rho_{ice} = 917 \frac{kg}{m^3}$ where the total ice shell mass is given by 
\begin{equation}
M_{ice} = V_{shell} * \rho_{ice} = \frac{4}{3}\pi \rho_{ice} (R_p^3 - (R_p - z)^3)
\end{equation}
The gravitational acceleration at the surface-atmosphere interface is derived by taking the planetary mass $M_p$ and subtracting off the mass of the ice shell, as well as the planetary radius minus the ice shell thickness 
\begin{equation}
g = G\frac{M_p - M_{ice}}{(R_p -z)^2}
\end{equation}
The surface pressure of the atmosphere (in bars) as a function of the thickness of the ice shell being sublimated is given by  
\begin{equation}
P_s(z) = \frac{M_{ice} * g}{\left[4 \pi (R_p - z)^2\right]} * 1\mathrm{e}{-5}
\end{equation}
where $R_{Europa}$ = 1560.8 km, $M_{Europa}$ = 4.8e22 kg, and an ice shell thickness up to 22 km \citep[e.g.,][]{ice_shell_thickness} can be completely sublimated to water vapor. Figure \ref{fig:Timescales} shows the surface pressure evolution. As more of the ice is sublimated, the surface pressure grows to nonphysical values ($\geq$ 100 bars, see Figure \ref{fig:Timescales}, top row, first panel).

Following the formalism of \citet{Ganymede_Europa_Migration}, we define this system as a Clausius-Clapeyron atmosphere which will limit the amount of the ice shell that can be sublimated. Clausius-Clapeyron interfaces occur where both the atmosphere and surface are the same molecule (in this case, water), the atmosphere being the gaseous state (water vapor) and the surface being the condensed state (liquid water or ice). The surface pressure for Clausius-Clapeyron interfaces is bounded by the saturation vapor pressure where, in equilibrium, the pressure a vapor imparts on its condensed phase cannot exceed the saturation vapor pressure. We utilize the Arden-Buck equation \citep{Buck1981} to compute the saturation vapor pressure: 
\begin{equation}
P_s(T)=
    \begin{cases}
        6.1121\mathrm{e}{-3} \: exp(T_i) & \text{if } Liquid \\
        \:\:\: T_i = (18.678-\frac{T-275.15}{234.15})(\frac{T-275.15}{T-18.01})\\
        6.1115\mathrm{e}{-3} \: exp(T_i) & \text{if } Ice\\
        \:\:\: T_i = (23.036-\frac{T-275.15}{333.7})(\frac{T-275.15}{T+4.67}) \; .
    \end{cases}
\end{equation}
As seen in Figure \ref{fig:Timescales} (top row, second panel), for an ice-vapor interface, \nt{for the maximum temperature of 220K (315K) found in our simulations in \S \ref{sec:Melting}, the surface pressure never exceeds 2e-5 (0.10 bars), even in the most extreme temperatures found in our simulations in \S \ref{sec:Melting}.} We keep the remaining analysis agnostic of the condensed phase of water on the surface. The saturation vapor pressure places an upper bound that regulates the amount of sublimated ice at any given point in time. Any time the pressure exceeds the saturation vapor pressure due to excess sublimation, the water vapor will condense back into a liquid or solid state. However, any water vapor that is lost to mass loss or photolysis can lower the surface pressure, and new ice can be sublimated. 

\subsection{Radiative Timescale and Greenhouse Effect}
The radiative timescale gives an estimate of how long it takes for a parcel of gas in the atmosphere to return to an equilibrium state once perturbed. The radiative timescale for Earth is on the order of weeks-months \citep{radiative_timescale_earth}, which is greater than the average time a parcel of gas spends on the night side of the planet. This is why the atmosphere on the night side of Earth retains heat. Following the formalism of \citet{Seager_Atmopsheres_Rad_Timescale}, the radiative timescale is given as
\begin{equation}
\tau_{rad}(T) \sim \frac{\Delta P (T)}{g} \frac{c_p}{4 \sigma_R T^3}
\label{eq:rad_timescale}
\end{equation}
where $\Delta P$ is the change in pressure from the surface to the top of the atmosphere, $c_p$ = 1.996e3 $\frac{J}{kg \: K}$ is the isobaric specific heat capacity, and $\sigma_R$ = 5.67e-8 $\frac{W}{m^2 \: K^4}$ is the radiation constant. The radiative timescale with the Clausius-Claperyon bounded surface pressure is shown in Figure \ref{fig:Timescales} (top row, third panel). The radiative timescale here grows large after $\sim$250K, meaning that water vapor could retain heat longer than the period of Europa's orbit ($\sim$85 hours). It is possible that equilibrium at the surface-atmosphere interface would be perturbed if a runaway effect occurs where more heat is deposited into the system with each subsequent orbit of Europa. \citet{Cosmic_Shoreline} find that a greenhouse runaway state would be attained if Europa receives Earth-like instellation. \citet{Ganymede_Europa_Migration} find that greenhouse effects are not significant for purely water-vapor atmospheres until an instellation threshold is crossed and that once crossed, water can heat up enough to become supercritical \citep{Supercritical_water_layer_inflation}. Once supercritical, the surface pressure can exceed the saturation vapor pressure, and sublimation can occur unimpeded. We consider both a bounded and unbounded surface pressure when computing hydrodynamic mass loss rates in the following section. 

\subsection{Hydrodynamic Escape}

Hydrodynamic escape is a pressure-driven thermal escape in which the surface pressure pushes up parcels of air past the sound speed, which then escapes en-masse into space. This form of atmospheric escape dominates in the small body regime \citep{Ganymede_Europa_Migration} where outward surface pressure dominates over gravity. By only considering gravity and surface pressure, hydrodynamic escape mass loss is given numerically by  
\begin{equation}
\dot{M}(T) = 4 \pi \rho_s(T) u_s(T) r_s^2
\label{eq:Hydro_Mass_Loss}
\end{equation}
where $M_p$ is the planetary mass, $G$ is the gravitational constant, $r_s$ is the surface radius, $\rho_s$ is the atmospheric surface density given by 
\begin{equation}
\rho_s(T) = \frac{P_s(T)}{u_0^2(T)}
\end{equation}
where $P_s$ is taken as the saturation vapor pressure, and $u_0$ is the isothermal sound speed given by 
\begin{equation}
u_0(T) = \frac{kT}{m} 
\end{equation}
where $k$ = 1.38e-23 $\frac{m^2 kg}{s^2 K}$ is the Boltzmann constant and $m$ = 2.989e26 kg is the mean molecular weight of water. 
$u_s$ is the outward radial flow velocity from the surface to the sonic level given by 
\begin{equation}
u_s(T) = u_0(T) \frac{r_c^2(T)}{r_s^2}\exp\left[\frac{-1}{2}-\frac{G}{u_0^2(T)}(\frac{3}{4 \pi \rho_m})^{-\frac{1}{3}}M_p^{\frac{2}{3}}\right]
\end{equation}
where  $\rho_m$ = 2500 $\frac{kg}{m^3}$ is the moon's mean density and $r_c$ is the radius at which gas reaches the isothermal sound speed
\begin{equation}
r_c(T) = \frac{GM_p}{2 u_0^2(T)} \; .
\end{equation}

Using this methodology and the saturation vapor pressure, the maximum mass loss rate due to hydrodynamic escape \nt{for the maximum temperature of 220K (315K) in the \S \ref{sec:Melting} simulations is 7e-6 kg/s (0.002 kg/s)} (see Figure \ref{fig:Timescales}, bottom row, first panel). Utilizing the unbounded surface pressure the maximum mass loss rate is of order magnitude 10 kg/s. Assuming that Europa's bulk mass is 5\% water, it would take \nt{~1e9 Gyr (3e6 Gyr)} to lose all of its surface water to hydrodynamic escape \nt{assuming the max temperature from our simulations} (or 750 Gyr with unbounded surface pressure). It is not expected that hydrodynamic escape would play a substantial role due to the bounded surface pressure combating the low gravity of the planet. 

\subsection{Jeans Escape}

When molecules in a parcel of gas are at a global temperature their velocity distribution is given by the Maxwell-Boltzmann distribution. At the tail end of the distribution, some molecules are above the escape velocity and detach from a planet's atmosphere and into space. The flux of escaping particles is given by
\begin{equation}
\Phi_{esc}(T) = \frac{n v_0(T)}{2 \sqrt{\pi}} \left(\frac{v_{esc}^2}{v_0^2(T)}+1\right) e^{-\frac{v_{esc}^2}{v_0^2(T)}}
\label{eq:Jeans_flux}
\end{equation}
where $n$ is the number density of particles, $v_0$ is the most probable velocity (the peak of the Maxwell-Boltzmann distribution), and is given by 
\begin{equation}
v_0(T) = \sqrt{\frac{2kT}{m}}
\end{equation}
and $v_{esc}$ is the escape velocity at altitude $y$ is given by
\begin{equation}
v_{esc} = \sqrt{\frac{2GM_p}{R_p}} \; .
\end{equation}

As seen in Figure \ref{fig:Timescales} (bottom row, second and third panels), the worst case mass loss rates \nt{for the maximum temperature of 220K (315K) in the \S \ref{sec:Melting} simulations is $\sim$0.05 kg/s (1.6e4 kg/s) for H$_2$O and 3.8e6 kg/s (5.3e9 kg/s) for H$_2$ gas. This means that it would take 1.6e5 Gyr (0.45 Gyr) years for Europa to lose all of its surface water assuming the maximum temperature from our simulations and that no water gets photolyzed into hydrogen gas.} Any H$_2$ gas can be assumed to be lost rapidly. This means that Jeans escape processes dominate over hydrodynamic escape. The low gravity of Europa results in a low escape velocity. However, water is a heavy molecule and, therefore, escapes less readily than hydrogen gas. 

\subsection{Photolysis and Clouds}

\nt{It is conjectured that} Venus lost a whole ocean's worth of water in the early history of the solar system \cite[e.g.,][]{Venus_Losing_Water} due to the ionization of H$_2$O into H$_2$ gas \nt{(though there are recent developments that point towards Venus potentially forming dry \citep[e.g.,][]{Constantinou2025})}. We utilized the photochemical 1D model \texttt{VULCAN} \citep{Tsai2017, Tsai2021} to get rough estimates on the effect of photolysis in a low-gravity, water-rich planet. Using the Sun's current spectra (incident at 1 AU), we initialize a Europan-liken object with an H$_2$O log volume mixing ratio of -2. Preliminary results show that the atmosphere will remain water-rich with H$_2$O remaining as a gas, being condensed into a low-density cloud that spans a large pressure range, or photolyzing into H$_2$, O$_2$, and O$_3$. Water being condensed into a cloud would help combat water loss due to photolysis, however photolysis does occur to a certain extent and splits the water molecule into easily lost hydrogen gas. A more complete model tracking the evolution between photolysis and atmospheric escape is beyond the scope of this work but could be important given that H$_2$ is lost extremely fast. Additionally, ionization rates depend on the UV properties of the red giant and would need to be modeled accurately over the entire evolution of the red giant branch phase. Notably, \citet{Lorenz1997} find that the UV flux of the Sun's future red giant phase will decrease with time, which in turn decreases haze production and UV-driven mass loss via photolysis during the red giant branch.



\subsection{Results and Discussion}

 We explore first-order approximations and effects of surface pressure evolution, the radiative timescale, the greenhouse effect, atmospheric escape, photolysis, and cloud formation on the evolution and stability of a tenuous Europan atmosphere \nt{as it varies with surface/atmosphere temperatures}. Due to the Clausius-Claperyon interface, we show that the amount of water that can be sublimated is bound by the saturation vapor pressure, keeping the surface pressure relatively low. We show that the radiative timescale of water vapor grows longer than the period of Europa's orbit at temperatures larger than $\sim$250K, meaning that heat is retained with each subsequent orbit. This could, in conjunction with the greenhouse effect, cause a runaway where the surface pressure can exceed the saturation vapor pressure due to supercritical surface water. Taking both a bounded and unbounded surface pressure into account, we compute atmospheric mass loss rates for hydrodynamic and Jeans escape. Jeans escape dominates over hydrodynamic escape, with water vapor taking 0.45-1.6e5 Gyr to be lost completely \nt{assuming the maximum temperatures from our simulations in \S \ref{sec:Melting}}. However, any water vapor that is photolyzed into hydrogen gas can be lost near-instantaneously, \nt{driving a more rapid loss water}. We explore the possibility of photolysis depleting the water vapor utilizing \texttt{VULCAN} and find that water vapor can be condensed into a low-altitude cloud, which would help combat water vapor loss due to photoionization into hydrogen gas.

All mass loss rates shown in Figure \ref{fig:Timescales} have a sharp upturn that occurs near $\sim$250K. \nt{In \S \ref{sec:Melting} we modeled the 2D surface temperatures of Europa when the Jupiter-Europa system first enters the red giant branch habitable zone (12.25 Gyr, S$_{\mathrm{eff}}$ = 0.32, Jupiter at 2 AU \citet{Cahoy2010} albedo), and when it recieves Earth-like instellation (12.45 Gyr, S$_{\mathrm{eff}}$ = 1.0, Jupiter at 0.8 AU \citet{Cahoy2010} albedo). In the first case, the temperatures in both hemispheres never reach high enough temperatures to have substantial mass loss (with a max temperature of 220K in the sub-Jovian equatorial region, see Figure \ref{fig:2AU-temperatures}). When Europa receives Earth-like instellation, the max temperature of both hemispheres do exceed 250K (see Figure \ref{fig:08AU-temperatures}). However, the orbit of Europa is relatively fast, and oscillations in temperature due to varying stellar and reflected light, and the Jupiter eclipse could effectively quench atmospheric loss processes. Both hemispheres heat and cool rapidly, and the sub-Jovian hemisphere never reaches its peak temperature due to the Jupiter eclipse.}

Additionally, as more atmospheric vapor builds up it is possible that the atmosphere would develop dynamics such as jets that homogenize any instantaneous temperature fluxes and combat atmospheric loss. Future work should consider implementing a coupled atmospheric-escape, UV photolysis, red giant branch stellar properties, and greenhouse effect model with Clausius-Clapeyron type interface to more accurately model this system. \nt{While our analysis is only considering two snapshots of the Jupiter-Europa system while its the red giant branch habitable zone, we note the incident flux on Europa's surface will evolve and increase throughout the entire red giant branch. While initially stable when the system first enters the habitable zone at 12.25 Gyr, at 12.45 Gyr the system begins to become more unstable as temperatures exceed 250K for a portion of the orbit. Past 12.45 Gyr, the Jupiter-Europa system will eventually receive S$_{\mathrm{eff}}$ = 1.11 (1524 W/m$^2$, at about 12.475 Gyr) where afterwards a runaway greenhouse is expected to occur. The Jupiter-Europa system will spend about 0.5 Gyr outside the habitable zone, after which the star will evolve on the asymptotic giant branch and the Jupiter-Europa system will once again be in the habitable zone \citep{Jupiter_in_HZ}.}

\nt{Taking the above into consideration, we can extrapolate our temperature dependent mass loss rates at the maximum temperatures in our simulations and conclude that Europa to first-order will keep its water for at least 0.2 Gyr as the incident flux evolves from S$_{\mathrm{eff}}$ = 0.32 to S$_{\mathrm{eff}}$ = 1.0, but that the fate of it's surface water afterwards will require more complete models that capture atmospheric dynamics. We note that its possible that even with increasing incident flux, second-order effects such as homogenizing jets, cloud formation, and latitude-longitude variations in heating and cooling of vapor can combat water loss over longer timescales. In particular, we show that in our simulations that only the equatioral regions of Europa reach the maximum temperatures of note (Figure \ref{fig:08AU-temperatures}), and that they cool relatively fast or are quenched by the eclipse of Jupiter.}

\section{Potential Detectability as an Exomoon}\label{sec:Exomoon}




\begin{figure*}
     \centering
         \includegraphics[width=1.0\linewidth]{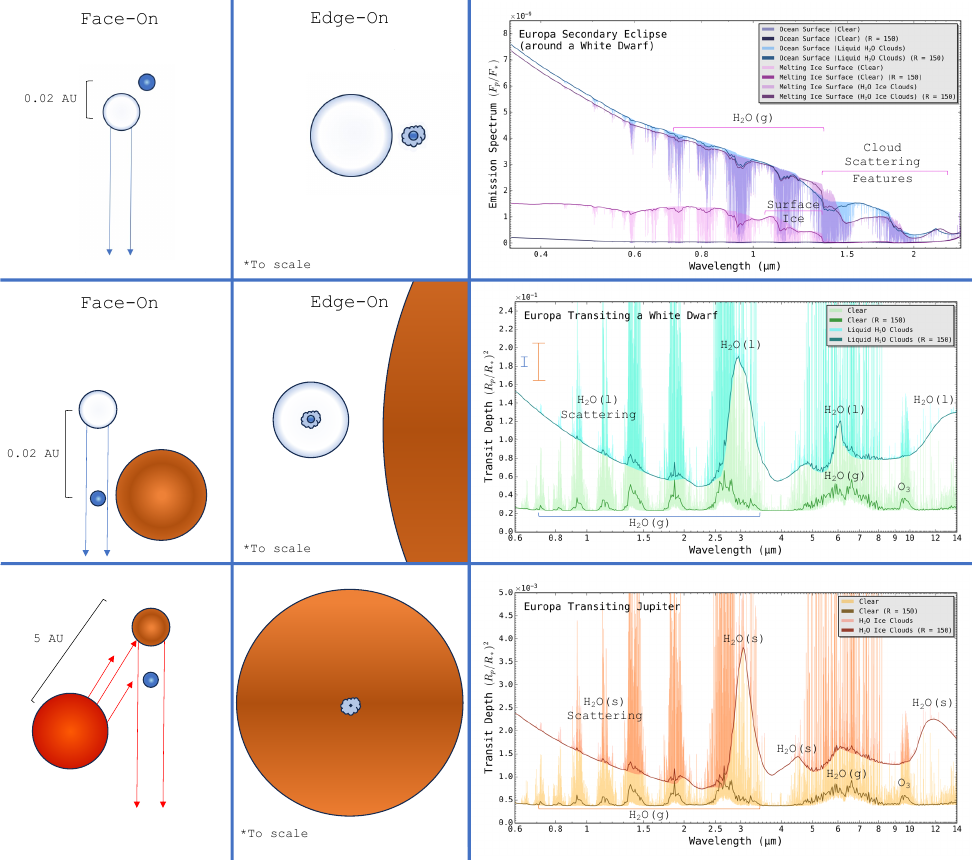}
     \caption{\nt{Potential methods by which to measure spectra of a sublimating Europan-like exomoon, with the atmosphere initialized from a photochemical model and included water, ozone, molecular oxygen, and molecular hydrogen. Spectra were generated with \texttt{POSEIDON} \citep{MacDonaldMadhusudhan2017, MacDonald2023, Mullens2024}. Cloudy models display the Mie scattering and absorptive properties of liquid water (H$_2$O (l)) and ice 1h (H$_2$O (s)). The figure displays the relative sizes of the white dwarf, moon, and host planet to scale, but not the orbital scale. Top: Dynamical models \citep{Payne2016} have shown that moons are likely liberated from their host planets and can occupy short distance orbits to white dwarfs. These spectra (0.35-2.45 \textmu m) demonstrate potential reflection+emission spectra of a liberated icy exomoon. We utilize two different surfaces (USGS Spectral Library's melting ice and ocean \citep{Kokaly2017,GoodisGordin2024}), as well as display clear and cloudy models. Middle: Gas giants, such as WD1856+534b, have been discovered transiting in the habitable zone (0.02 AU) of white dwarfs \citep{WD1856b_Discovery}. If dynamically stable, transiting giant planets can have their moon transit as well, producing high SNR transit signals due to the size ratio between the moon and white dwarf (0.6-14 \textmu m). Representative error bars for JWST NIRSpec Prism (blue) and MIRI LRS (red) are shown. Bottom: Gas giants hosting moons during the red-giant branch reflect light from their host star, as well as emit thermal emission (0.6-14 \textmu m). Future direct-imaging instruments, such as Roman Space Telescope and HWO, might be able to capture exomoon transits around Jupiter analogs and emission signals from Europan-analogs around white dwarfs, but these are both currently inaccessible to JWST.}
        }
     \label{fig:spectra}
\end{figure*}

 Given the modeling done in \S \ref{sec:Atmosphere}, we assume that Europa around Jupiter would retain some sort of water-rich atmosphere during portions of the \nt{red giant branch}. \nt{While our models demonstrate that Europa would likely lose a substantial amount of its water during the later stages of the red giant branch, we assume that our Europan-like analog could retain its surface water throughout the red giant branch (either due to a larger initial orbital distance from their host star, or other second-order mechanisms explored in \S \ref{sec:Discussion}}). We propose \nt{three} ways by which a water-rich Europan-like exomoon exposed to \nt{habitable zone} instellation could be detected remotely.
 
 \nt{The first method is the measuring the secondary-eclipse spectrum of a Europa-analog that has been liberated from its host-planet and occupies the habitable zone of a white dwarf (dynamical models have shown that this is a possibility, see \S \ref{sec:Discussion} and \citet{Payne2016}). The \nt{second} method is similar in concept, and would be a transit caused by the moon around a white dwarf, either during or right before/after the transit of the host planet.} \nt{In this scenario, the host-planet and its moon has survived the evolution of the star through the red giant branch and asymptotic giant branch, and later migrated to the habitable zone of the resultant white dwarf. This is analogous to the giant planet WD1856+534b \citep{WD1856b_Discovery}, which survived the post-main sequence evolution of its host star and now orbits within 0.02 AU (8 minute transit duration) of the resultant white dwarf.} Previous studies have investigated the detectability of exomoons during their host planet's transit \citep[e.g.,][]{Harada2023} and have found that moons can contaminate transmission signals during the host planet's primary transit at the $\sim$10 ppm level. However, with the relatively small size of white dwarfs, it is therefore entirely feasible that for a given orbital geometry, an exomoon around a Jupiter-sized exoplanet could have its own transit signal independent of its host planet. White dwarfs have on average a diameter of $\sim$7000 km and the average orbital distance of Europa from Jupiter is 671,000 km, or about $\sim$100x larger than the diameter of a nominal white dwarf.  Given the size ratio of a white dwarf to the moon, the signal would be detectable, and previous work has shown the benefits of measuring rocky planet transmission spectra around white dwarfs \nt{with observatories such as JWST, despite white dwarfs emitting mostly in the UV} \citep{White_Dwarf_Jupiter}. However, the system geometry would have to be near perfect, and a secondary transit of this variety would only be resolved with long-baseline observations that would get lucky enough to have a moon situated on the correct side of the host planet. \nt{On the other hand, a liberated exomoon orbiting at 0.02 AU could also transit, though no transiting terrestrial planet around a white dwarf has been discovered as of yet.}

The \nt{third} method would be using the \nt{host planet's reflected light and intrinsic thermal emission} during the \nt{red giant branch} to conduct transit measurements across the host planet's reflecting surface. Jupiter itself is extremely reflective \nt{when it has clouds in its atmosphere (see 5 AU and 2 AU \citet{Cahoy2010} geometric albedo models)}, and combined with a higher incident flux hitting Jupiter during the red giant branch and the orbital distance remaining large enough to feasibly use a coronograph, one can view the geometry of the system near a secondary eclipse and use the reflected light off of a host-planet to conduct transit measurements of an exomoon. \nt{Future missions, such as Roman Space Telescope and the Habitable Worlds Observatory, will be equipped with coronographs and be potentially sensitive to Jupiter-analog emission signals.} Additionally, given Europa's short orbit compared to Jupiter's long orbit, the window of opportunity to observe this would last on the order of a few weeks allowing an observer to build up a sufficient signal to detect and characterize a `transiting' exomoon. Given Jupiter's relative size to Europa, the transit signal would be smaller than the white dwarf case but could be buffered with repeated exomoon transits. Given Europa's reflectivity it is possible that `dayside reflection contamination' could tamper the spectra. Since observed emitted reflection spectra is $= F_s(\lambda) \: A_g(\lambda) \: (R_{p}/a_{p} d)^2 $ where $R_p$ is the reflecting planet's radius, $a_p$ is the orbital distance from the star, and $d$ is the distance to the system, $F_s$ is the stellar flux and $A_g$ is the geometric albedo, the signal from Europa's dayside would be expected to be minimal compared to Jupiter's signal due to their radius ratio. Geometries for both of these systems can be seen in Figure \ref{fig:spectra}. 

 We model potential \nt{emission and} transmission spectra of a water-rich Europan-liken exomoon in \nt{all three} of these geometries with the open-source atmospheric retrieval code POSEIDON \citep{MacDonaldMadhusudhan2017, MacDonald2023}. Taking isochemical averages from the \texttt{VULCAN} model in \S \ref{sec:Atmosphere}, we initialize an atmosphere with an isothermal 300K pressure-temperature profile and with H$_2$O (log volume mixing ratio = -2), O3 (-6), O2 (-3), and bulk H$_2$ (no He). We model spectra with and without water clouds. We model the water clouds utilizing the Mie-scattering algorithm implemented in POSEIDON and presented in \citet{Grant2023} and \citet{Mullens2024}. The water ice 1h refractive indices used to generate the aerosol properties are from \citet{Warren1984} and the liquid water refractive indices are from \citet{Hale1973}. We initialize the water clouds with a mean particle size of log r$_m$ = -1.5 (\textmu m), a low log volume mixing ratio (-11), and let the cloud span the entire pressure range. We additionally take a hard surface at the saturation vapor pressure of water at this temperature (0.061 bars). \nt{For the emission spectra, we allow the surface to emit and reflect radiation depending on its albedo (which is currently a beta feature in \texttt{POSEIDON}, Mullens et al 2024, in prep) and test both a melting ice and ocean surface from the USGS Spectral Library \citep{Kokaly2017,GoodisGordin2024}. The results are seen in  Figure \ref{fig:spectra}.} 

\nt{The emission spectra are generated with both reflection and thermal scattering enabled. If clear, the emission spectra (0.35-2.45 \textmu m) are sensitive to H$_2$O bands and the surface composition (with a melting ice surface having a higher albedo). If cloudy, the water clouds have a high albedo and produce a brighter spectra with H$_2$O bands imprinted on the spectra.} The resultant clear transmission spectra for the two transit cases are identical save the magnitude of the transit depth. In the clear models one can clearly detect strong H$_2$O bands as well as a weak O$_3$ feature around 10 \textmu m. In cloudy models, we clearly detect an amplified scattering slope in the short wavelengths that mutes any spectral features up to $\sim$1.4 \textmu m. Liquid water and ice 1h have a large spectral features at 3 \textmu m that would be easily detectable, with additional features at 6 \textmu m (liquid water) and 4.5 \textmu m (ice 1h). Cloudy models obscure any ozone features. It is worth noting that if water ice has a higher mixing ratio than $\sim -10$ and is homogeneous throughout the atmosphere, it will completely wipe out the spectra and make the characterization of the moon inaccessible. However, detection will still be feasible. 

It is beyond the scope of this work to produce and run retrievals on synthetic data produced in our proposed observing geometries. Utilizing white dwarfs to study rocky planet atmospheres has been explored in extant work. \cite{White_Dwarf_Jupiter} simulated JWST data and showed that transiting rocky planets within \nt{the habitable zones (0.005-0.02 AU) \citep{Kozakis2018}} of white dwarfs can have H$_2$O abundances detected strongly and O$_3$ abundances detected weakly within 5 transits. Our \nt{third} observing method is much more theoretical in its premise. However, given Jupiter's high reflectivity and large radius, one can imagine that the signal that can be achieved during the red giant branch would be high enough to make this a practical approach \nt{with future observatories such as Roman Space Telescope and Habitable Worlds Observatory}.

\nt{We utilize the JWST ETC and Pandexo \citep{batalha2017_pandexo} in order to compute SNR estimates for a Europan-like analog transiting a white dwarf (we use WD1856+534 as a proxy star, and compute SNR for 10 transits). With JWST instruments the secondary eclipse spectra will be inaccessible due to the low emission signal, which is minimal due to Europa's small radius (0.245 R$_\mathrm{E}$) and low temperature. It will, however, be within HWO/LUVOIR's (0.2-2.5 \textmu m) expected contrast ratio \citep{Juanola-Parramon2022}. The transmission spectra is much more feasible to measure due to its dependence on the radius ratio between the Europan-analog and the white dwarf, with representative error bars for JWST NIRSpec PRISM (0.6-5 \textmu m, blue) and MIRI LRS (5-10 \textmu m, red) in Figure \ref{fig:spectra}. A Jupiter-analog at 5 AU is well within the inner working angle of of Roman Space Telescope, where the spectroscopic channel will only be able to access 0.6-0.8 \textmu m \citep{Kasdin2020}.} Future work will need to test whether or not it is feasible to utilize a coronograph to block out the light from the red giant sufficiently enough to detect a reflecting planet in this geometry, \nt{and if a time-series observation would be capable of detecting exomoon transits.}

\section{Discussion and Further Considerations}\label{sec:Discussion}

We have shown that Europa's surface sublimates readily \nt{when in the habitable zone of the red giant branch} and that the anti-Jovian, sub-Jovian, north, and south hemispheres display interesting asymmetric behaviors with diurnal and seasonal variations. We also showed that atmospheric escape is subdued compared to expectations for a low-mass exomoon due to the unique surface-atmosphere interface formed by sublimation, \nt{and that Europa can retain its surface water for at least 0.2 Gyr in the red giant branch habitable zone.} We now seek to place this system in the context of the red giant branch and white dwarf phases.

The first hurdle of the red giant branch would be the stellar winds which could ablate the surface layers of Europa. It is unlikely that Jupiter's atmosphere would be completely eroded by stellar winds but it is possible that its composition would change via accretion \citep{Veras_Post_MS_Review}. However, stellar winds could potentially be important for orbiting moons. Jupiter's current magnetic field is large and strong enough to deflect stellar winds. Taking into account magnetic field evolution  \citet{Jupiter_Magnetic_Field_Post_MS} show to first order Jupiter's magnetopause could be reduced in size to one Jupiter radius during the post-main sequence. This implies that ablation could be significant for Europa's survivability since Europa orbits at $\sim$10 Jupiter radii \citep{Europa_orbital_data}. However, \citet{Jupiter_Magnetic_Field_Post_MS} don't take into account Jupiter's magnetic field inflation due to thermal and plasma pressure. Additionally, Jupiter's current magnetic field has a minimum radius pointing towards the stellar winds and is larger pointing away, implying that Europa could partially be in Jupiter's magnetic field for a portion of its orbit and experience oscillatory effects from stellar winds. We additionally neglect the effect of Jupiter's magnetic field stripping the atmosphere (as it does currently with Io's tenuous atmosphere), which could accelerate any atmospheric loss of photochemically produced ions \citep{Bagenal2020}. 

Another potential issue is that increased instellation can inflate Jupiter's size. It is possible that Jupiter's outer layers would heat up enough to `puff' up. Extreme cases of irradiation have caused planets half the mass of Jupiter to expand only up to a 1/3 of the size that they are expected to be \citep{inflated_jupiters}. The increase in temperature from its current temperature (165K) to the Earth-like instellation would not change the scale height of Jupiter's atmosphere by much since Jupiter's high gravity dominates. Therefore, Europa and most larger moons would remain comfortably safe orbiting their gas giant. Potentially there is a safe zone where an icy moon would be within the magnetic field of its host planet but not close enough to be within the planet's tidal disruption zone or extended atmosphere. 


This work additionally neglects to model the tidal evolution of Europa during the red giant branch and in the habitable zone of a white dwarf. \citet{Zollinger2017} show that exomoons are not stable within the habitable zone of dwarf stars due to intense tidal heating. We defer any investigations to the tidal heating evolution of Europa and its implications with respect to interior melting, surface evolution, and orbital instability to future work. One potentially interesting avenue is the melting of the moon's core, thereby making it easier for the moon to dissipate tidal heating. Another is an increase in interior heat that could cause the inner part of the ice shell to melt, thereby providing an interesting avenue by which to have surface liquid water. Being a moon in the habitable zone of a white dwarf could provide a mechanism by which a terrestrial world retains a melted interior and, therefore, a geologic cycle that has been shown to be potentially necessary for life while orbiting a white dwarf that remains stable on age-of-the-universe timescales \citep{Kozakis2018}. 

\citet{Payne2016} explore the dynamical evolution of exomoons during the post-main sequence. They find that during the red giant branch and subsequent asymptotic giant branch phase that moons around giant planets become more stable due to the host planet's Hill radius growing (soley due to stellar mass loss). However, exomoons can be subsequently liberated from their host planet in the white dwarf phase. Close-planet-planet encounters (gravitational scattering) can frequently liberate exomoons and cause them to orbit at large distances from the white dwarf. Another possibility is the host planet being scattered into an eccentric orbit and the moon being liberated during periapse passage, which occurs less frequently but causes exomoons to occupy short distance orbits from the white dwarf. Short orbit, liberated, tidally disrupted exomoons that form initially from irradiated ice found in rings systems within giant planet radiation belts have been invoked to explain excess beryllium and lithium found in polluted white dwarf spectra \citep{Doyle2021}, though this explanation has been brought under scrutiny \citep{Kaiser2024}. \citet{Payne2016} find that some liberated exomoons remain in a stable, close-in orbit around the white dwarf (assuming no further scattering events) which could allow for a liberated exo-Europa to end up in the habitable zone of the white dwarf without its host planet, making it amenable to characterization via transmission or emission spectroscopy (a scenario features in Figure \ref{fig:spectra}). Without its host-planet, the exomoon would instead be tidally locked to the white dwarf leading to a different model of incident irradiation than explored in this work; \citet{Shields2024} find that tidally locked planets in the habitable zone of white dwarfs have homogeneous, warmer atmospheres when compared to main-sequence worlds due to strong jets that form due to a fast rotation and orbital period. We defer the exploration of surface evolution of such a tidally locked liberated exomoon to future work. 

Though the entirety of this work focuses on and extrapolates the Jupiter-Europa system, the methodology presented can be applied to other solar system satellites or exomoon systems. There exists an entire phase space to explore with both the moon and host-planet properties. Moons come in a variety of different masses, compositions, and distances from the host planet and could have widely different evolutionary paths in the post-main sequence. There additionally exists an entire phase space of planets with different sizes, distance from the host star, eccentricities, reflectivities, and obliquities that could affect the diurnal and yearly variations a moon experiences. In particular, a higher obliquity could impart interesting seasonal variations akin to the moons of Saturn \citep[e.g.,][]{Ashkenazy2016}. \citet{Lorenz1997} explored the evolution of Saturn's moon Titan under a red-giant star and found that the lower-UV flux of the red-giant lowers haze production and allows Titan to form liquid water-ammonia oceans.  Future work should explore this phase-space more completely to discover if there exists a subset of phase-space where a moon could have stable surface water and potentially an atmosphere for billions of years in the habitable zone. 

This large phase space can be extended to the detection methods described in \S \ref{sec:Exomoon}. A unique array of moon evolution pathways within a host-planet moon system could be detected around a single gas giant, analogous to the multi-transiting systems around red dwarf stars, allowing us to leverage atmospheric signals to probe the surface composition of now-ablated low-mass exomoons and use their atmospheres as laboratories for low-gravity, moderate-instellation atmospheric evolution. However, observations of moons using the techniques listed above could be hindered by the formation of a ring of condensed material around the host planet (analogous to Enceladus's E-Ring) due to atmospheric escape. It is a valid question as to how a ring of diffuse ice would affect any potential transit observations, either a secondary transit around a white dwarf or using the reflection method. As for host-planet considerations, sub-Neptune class planets could be particularly conducive to the `optical mirror' transit method. Sub-Neptune class planets are the most common type of planet, smaller (1-3.8 R$_\oplus$ [Earth radii]) in size and therefore have a better exomoon transit signal, and have been found to be more reflective on average than different classes of exoplanets \citep[e.g.,][]{Kempton-2023Nature}. Future work will be needed to accurately model exomoon atmospheres and exoplanet reflection during the red giant branch to explore this observational strategy in depth. 

\section{Conclusions}\label{sec:Conclusions}

We model the surface evolution of Europa when initially exposed to habitable-zone instellation \nt{in the red giant branch} in order to gain insights on the potential long-term stability of surface water. We include the diurnal flux variations due to Europa being tidally locked to Jupiter and the yearly flux variations from Jupiter's eccentric orbit and obliquity, as well as flux from stellar and reflected light, and thermal fluxes from Jupiter and Europa's interior. \nt{We explore two distinct snapshots of the Europa-Jupiter system in the red giant branch habitable zone: when it first enters the red giant branch habitable zone (12.25 Gyr, S$_{\mathrm{eff}}$ = 0.32, Jupiter at 2 AU \citet{Cahoy2010} albedo) and when it receives Earth-like instellation (12.45 Gyr, S$_{\mathrm{eff}}$ = 1.0, Jupiter at 0.8 AU \citet{Cahoy2010} albedo).} Implementing a Newtonian-cooling scheme \nt{with ice-vapor phase changes} to convert absorbed flux to a surface temperature, \nt{we find that when the Europa-Jovian system first enters the red giant branch habitable zone, the entire equatorial region of Europa sublimates while the mid-latitudes of the sub-Jovian hemisphere also sublimates due to increased flux from Jupiter's reflected light. We find that there is a significant diurnal asymmetry between the anti- and sub-Jovian hemispheres and a slight yearly asymmetry between the north and south hemispheres. In particular, vapor on the sub-Jovian hemisphere is kept warmer than the anti-Jovian hemisphere due to the reflected light from Jupiter, with Jupiter eclipsing the Sun keeping the temperature from reaching its maximum value. Northern latitudes in both hemispheres experience more drastic seasonal changes due to the peaks of the substellar latitude aligning with Jupiter's eccentric orbit. When the system receives Earth-like instellation, both hemispheres are substantially sublimated and have similar diurnal temperature variations due to the lower albedo of Jupiter once clouds disipate. }

Finding that the equatorial region readily sublimates, first-order approximations of mass loss find that in equilibrium, the surface pressure is bounded by the Clausius-Clapeyron nature of the ice-vapor interface. We also find that depending on the temperature, the radiative timescale can far exceed the diurnal period of Europa which could lead to a runaway effect. We find that hydrodynamic escape is negligible and that water and hydrogen gas jeans escape can be significant at certain temperatures. Above 250K, the radiative timescale and escape rates all experience a sharp increase. \nt{Taking the maximum temperatures from our simulations at 12.25 Gyr (220K) and 12.45 Gyr (315K), we find that as a lower bound water would take 0.45-1.6e5 Gyr to escape, allowing us to state that surface water would at least persist for 0.2 Gyr in the red giant branch habitable zone.}

It is unknown whether escape will happen locally or globally and if escape processes can be effectively quenched due to the oscillatory nature of absorbed diurnal flux and potential atmospheric dynamics that could homogenize atmospheric variations. A 1D photochemical model reveals that water vapor either photolyzes into oxygen and hydrogen gas, condenses into a low-density cloud that spans a large pressure range or remains as water vapor. The possible formation of a cloud and drop in UV flux as the sun evolves as a red giant could combat mass loss due to photolysis.

Using a 1D photochemical model to compute clear and cloudy spectra of a potential sublimating Europan-like exomoon, we present \nt{three} methods by which to detect and chararactize this system: \nt{a secondary eclipse of a liberated exomoon within the habitable zone of a white dwarf,} a secondary transit around a white dwarf, or utilizing reflected light off a host planet to measure a transit across a planet's reflecting surface. We acknowledge that the \nt{three} detection methods are theoretical and that future work needs to be done to weigh the merits of observing an exomoon utilizing them. 

Placing the system in the context of the post-MS, the effects of stellar winds, evolving flux, and tidal evolution could be significant for systems like Jupiter's as well as other planet-moon systems, such as the moons of Saturn that are different in composition and experience more extreme seasonal variations, and sub-Neptune exoplanets that could provide a more reasonable class of planet to consider the `optical mirror' transit method on. 

A majority of habitable terrestrial planets, including ours, will one day face their inevitable end in the red giant branch. However, icy moons in our solar system and around other stars could evolve into a habitable world around a white dwarf that remains stable for timescales longer than the age of the universe. It is possible that in a fiery dramatic end to one part of the system, life could find new beginnings.

\section{Data Availability}

\nttwo{The Zenodo repository for this paper (\href{https://doi.org/10.5281/zenodo.15375314}{Zenodo doi: 15375314}) contains all the materials needed to reproduce the results and figures of this work.}

\section*{Acknowledgements}
We thank the AbGradCon 2024 committee for supporting this project by picking it to be presented as in-person poster at the AbGradCon 2024 and Outer Planets Assessment Group 2024 joint poster session. E.M. acknowledges that this material is based upon work supported by the National Science Foundation Graduate Research Fellowship under Grant No.\ 2139899. \nt{E.M. thanks Ekaterina Landgren for help in jumpstarting this project, and Abby Boehm for the constant encouragement to publish this work. We thank the referee for a very constructive and thorough report of our manuscript that significantly improved the quality of this work, especially in regards to the inclusion of phase-changes in our simulation.}

\clearpage


\clearpage
\bibliographystyle{mnras}
\bibliography{sample.bib}

\appendix 

\renewcommand{\thefigure}{A\arabic{figure}}
\renewcommand{\thetable}{A\arabic{table}}

\begin{figure*}
     \centering
         \includegraphics[width=1.0\linewidth]{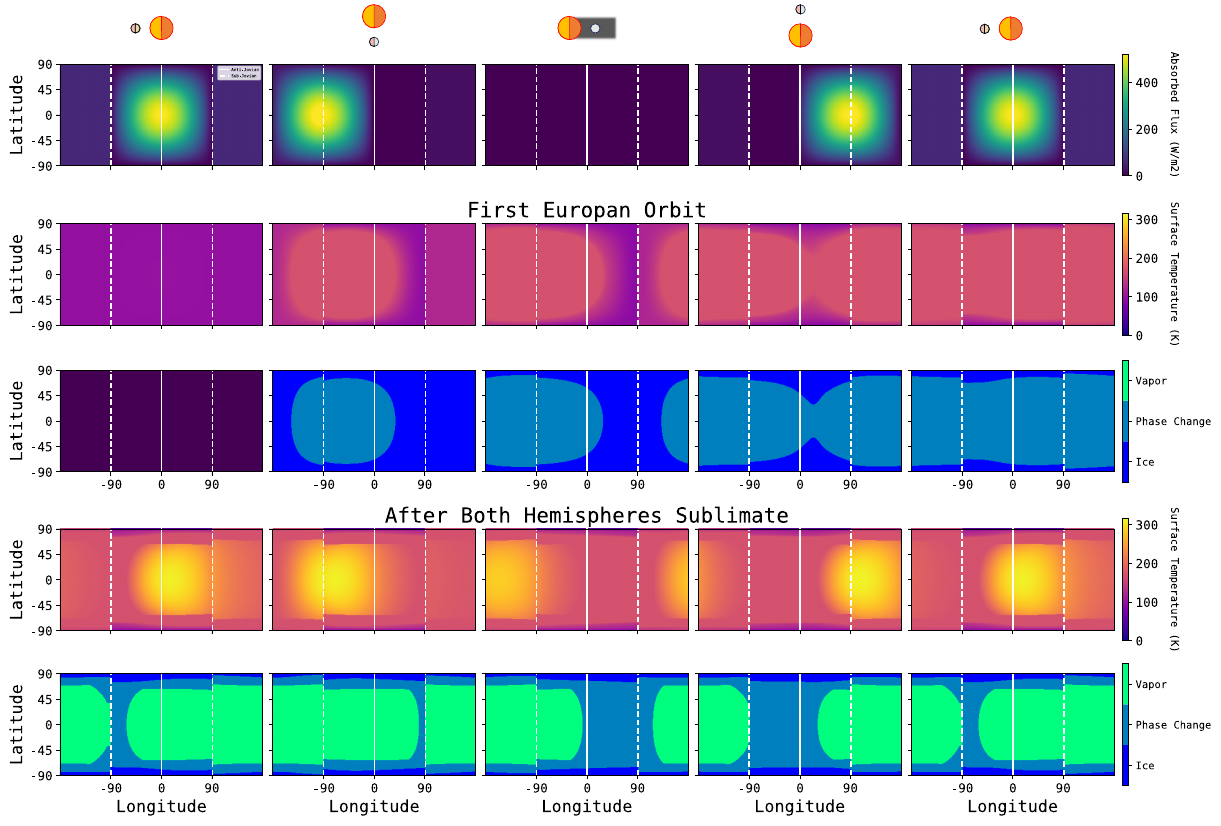}
         \caption{\nt{Same as Figure \ref{fig:2AU-Snapshots}, but for the Jupiter-Europa system when it receives Earth-like instellation (12.45 Gyr, S$_{\mathrm{eff}}$ = 1.0, Jupiter at 0.8 AU \citet{Cahoy2010} albedo). Due to the much lower albedo of Jupiter during this orbit, but the higher incident flux, the sub-Jovian and anti-Jovian hemispheres evolve nearly identically, both sublimating around the 2 Jupiter Years mark. Yearly and daily surface temperature variations after the simulation has reached steady state are shown in Figure \ref{fig:2AU-temperatures}.}}
        \label{fig:08AU-snapshots}
\end{figure*} 

\begin{figure*}
     \centering
         \includegraphics[width=1.0\linewidth]{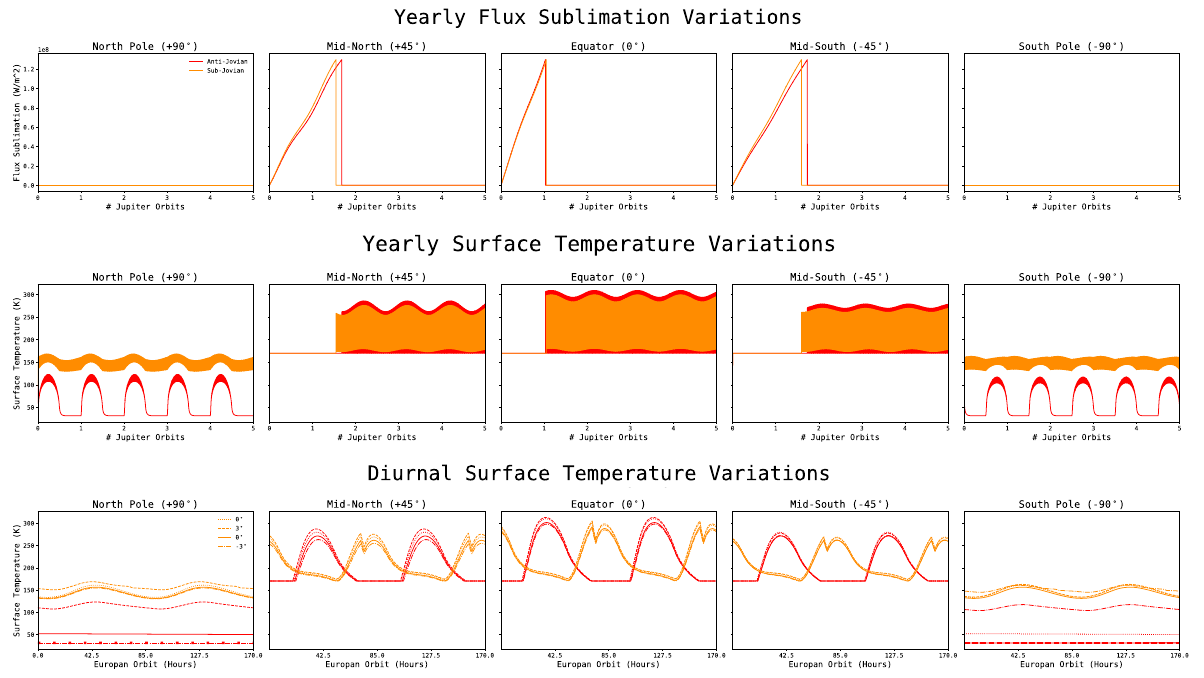}
         \caption{\nt{Same as Figure \ref{fig:2AU-temperatures}, but for the Jupiter-Europa system when it receives Earth-like instellation (12.45 Gyr, S$_{\mathrm{eff}}$ = 1.0, Jupiter at 0.8 AU \citet{Cahoy2010} albedo). Both the sub-Jovian and anti-Jovian mid-latitudes and equator undergo the phase transition to vapor in less than 2 Jupiter years, with both hemispheres being more symmetric due to the lower bond albedo of Jupiter at this point in the red giant branch. Due to Jupiter's eccentric orbit there are variations in the peak surface temperature where the northern latitudes display more extreme seasonal differences due to Europa experiencing northern summer during Jupiter's periapse. Contrasting the results of the simulation in the main text, the anti-Jovian hemisphere here reaches a peak temperature higher than the sub-Jovian hemisphere only due to cooling during the eclipse. The max surface temperature of the anti-Jovian (sub-Jovian) hemisphere is 314K (306K).}}
         \label{fig:08AU-temperatures}
\end{figure*} 

\bsp	
\label{lastpage}
\end{document}